# Engineering of Heterostructure $Pt/Co/AlO_x$ for the enhancement of Dyzaloshinskii-Moria interaction

Babu Ram Sankhi[a], Elena Echeverria[a], Hans T. Nembach[b,d], Justin M. Shaw[b], Soumya Mandal[c], Muhammet Annaorazov[a], Ritesh Sachan[c], David N. McIlroy[a], D. Meyers[a], Emrah Turgut[a]

[a] Department of Physics, Oklahoma State University, Stillwater, Oklahoma 74078-3072, USA

[b] National Institute of Standards and Technology, Boulder, Colorado 80305, USA

[c] Department of Mechanical and Aerospace Engineering, Oklahoma State University, Stillwater, Oklahoma 74078-3072, USA

[d] Department of Physics, University of Colorado Boulder, Boulder, CO 80309, USA

## Abstract

The interfacial Dyzaloshinskii-Moria interaction (DMI) helps to stabilize chiral domain walls and magnetic skyrmions, which will facilitate new magnetic memories and spintronics logic devices. The study of interfacial DMI in perpendicularly magnetized structurally asymmetric heavy metal (HM) / ferromagnetic (FM) multilayer systems is of high importance due to the formation of chiral magnetic textures in the presence of DMI. Here, we report the impact of the cobalt oxidation at the cobalt -aluminum oxide interface in $Pt/Co/AlO_x$ trilayer structure on the DMI by varying the post-growth annealing time and Aluminum thickness. For quantifying DMI, we employed magneto-optical imaging of asymmetric domain wall expansion, hysteresis loop shift, and spin-wave spectroscopy techniques. We further correlated the Cobalt oxidation with low-temperature Hall effect measurements and X-ray photoelectron spectroscopy. Our results emphasize the characterization of magnetic films for MRAM technologies semiconductor temperature process window, where magnetic interaction will be critical for device performance.

## 1. Introduction

The Dyzaloshinskii-Moria interaction (DMI)[1–3] is an antisymmetric exchange interaction between spins induced by broken inversion symmetry in magnetic materials either structurally or crystallographically. The DMI energy between any two neighboring spins $\boldsymbol{S_i}$ and $\boldsymbol{S_j}$ can be written as $H_{DMI} \propto (\boldsymbol{S_i} \times \boldsymbol{S_j})$, which causes non-collinear alignment of magnetic moments. The competition between DMI and Heisenberg interaction $\propto (\boldsymbol{S_i} \cdot \boldsymbol{S_j})$ leaves the spins in a chiral state rather than a ferromagnetic state, which leads to the formation of exquisite spin textures such as chiral domain walls (DWs)[4], skyrmions[5], and other topological spin textures[6,7].

These topological properties of chiral spin textures have received considerable attention due to their possible applications in non-volatile spintronics devices [5,8–10] such as racetrack memories [10,11]. In addition, they can be driven at a very high speed with a low-charge current using spin-orbit torques (SOT) and Zhang-Li spin-transfer torque [4,12–16]. In recent years, efforts were made to study current-induced chiral DW motion [4], skyrmion dynamics [8,17,18] as well as to study the role of DMI in controlling the size of a skyrmion [19].

An important class of material systems to optimize the DMI is the heavy metal (HM)/ferromagnetic (FM)/Oxide multilayer (ML) system with a perpendicular magnetic anisotropy (PMA)[20–24], where a strong spin-orbit interaction in the HM layer modifies the DMI. Moreover, the FM layer's oxidation alters the PMA, which has been extensively studied previously[25–27]. However, the effect of oxidation of the FM on DMI has not been understood thoroughly. Quantification of the DMI in these ML systems has great importance to accelerate the realization of spintronics applications based on chiral spin textures. In this article, we control and quantify the DMI in a $Pt/Co/AlO_x$ trilayer system by varying the Aluminum thickness, oxidation strength, annealing time, and the substrate. After growing the multilayer films with a magnetron sputtering technique, we use asymmetric domain wall (DW) expansion using polar Magneto-optical Kerr microscopy (MOKE) [28–31]. We fabricate micron-scale Hall-bar structures and determine magnetic anisotropy and DMI effect on hysteresis curves. As a third method, we employ Brillouin Light Spectroscopy techniques for the quantification of the DMI constant. Moreover, the enhancement of DMI is due to the different $CoO_x$ content at the interface as evidenced by results from the X-ray photoelectron spectroscopy (XPS) and low-temperature anomalous Hall effect (LT-AHE) measurement technique.

## 2. Experimental details

The multilayers (MLs) $Ta(10)/Pt(4)/Co(1.2)/AlO_x(t)/Pt(6.7)$ were grown on thermally oxidized silicon and sapphire substrates, where the numbers in the bracket represent their thickness in nm. The base pressure of the system is less than $10^{-6}$ Pa and the growth pressure are maintained at 267 mPa. In these MLs, Tantalum is used to have a smooth surface for the succeeding deposition, Platinum is a heavy metal having high spin-orbit coupling, and Aluminum was introduced to suffice the condition of structural inversion symmetry breaking for the enhancement of DMI. The aluminum layer was oxidized with Oxygen plasma for 15 seconds with the pressure maintained at 2.7 Pa and 44-Watt RF power. The top Pt layer acts as a capping layer to prevent further oxidation of the aluminum layer. We post-annealed the films at 350° C for varying time intervals in the growth chamber. Following the growth and overnight cooling, we characterize the un-patterned films using MOKE microscope and vibrating sample magnetometry (VSM) for saturation magnetization and magnetic anisotropy values. Next, we use the ferromagnetic resonance technique to measure the damping parameter and magnetic inhomogeneity, the spectroscopic g-factor, and anisotropy. For the hysteresis loop shift measurement, the samples were patterned in the 27-µm Hall bar devices and the current density applied across the devices for the DMI measurement was $\sim 10^{10}~A/m^2$ .The hall resistance was measured as a function of the out-off plane (OOP) magnetic field $H_z$. The interfacial DMIs are quantified by using the domain wall expansion method in the creep regime by using polar MOKE microscopy, hysteresis loop shift (HLS) method, and BLS measurement techniques. The layered structure of the stacks was ensured by JEOL JEM-2100 transmission electron microscopy (TEM) operating at 200 kV and different chemical environments of material in MLs were investigated by using XPS and LT-AHE measurement techniques. For XPS measurements, the surface composition was measured for as-grown and sputtered films to study the Co oxidation at the $Co/AlO_x$ interface. All samples were sputtered with $Ar^+$ at 6.65x10$^{-3}$ Pa. Sputter times were varied depending on the thickness of each sample.

## 3. Results and Discussion

### a. Magnetic characterization

The square hysteresis curves were measured by sweeping the $H_z$ field (Fig. 1(a)), which confirms the strong PMA nature of our samples. The saturation magnetizations ($M_s$) of the samples were derived from the magnetic hysteresis loops (presented in the supplementary material Fig.S1) measured by using VSM[32]. The anisotropy fields ($H_k$) were extracted from the magnetoresistance curve of the IP hysteresis loops [31] by applying a low DC of the order of $1\ mA$. The effective anisotropy fields ($K_{eff}$) depends on the $M_s$ and $H_k$, and calculated from the relation $K_{eff} = \frac{1}{2} \mu_0 M_s H_k$ [31,33,34]. The variation of areal magnetization ($M_s t$) and $K_{eff}$ both lie within the range of less than 25% of their average value for all the samples, as illustrated by Fig. 1(b). This variation of the magnetic parameters with annealing time is an indication of a change in cobalt concentration in some extent due to diffusion of oxygen at ferromagnet-metal interface.

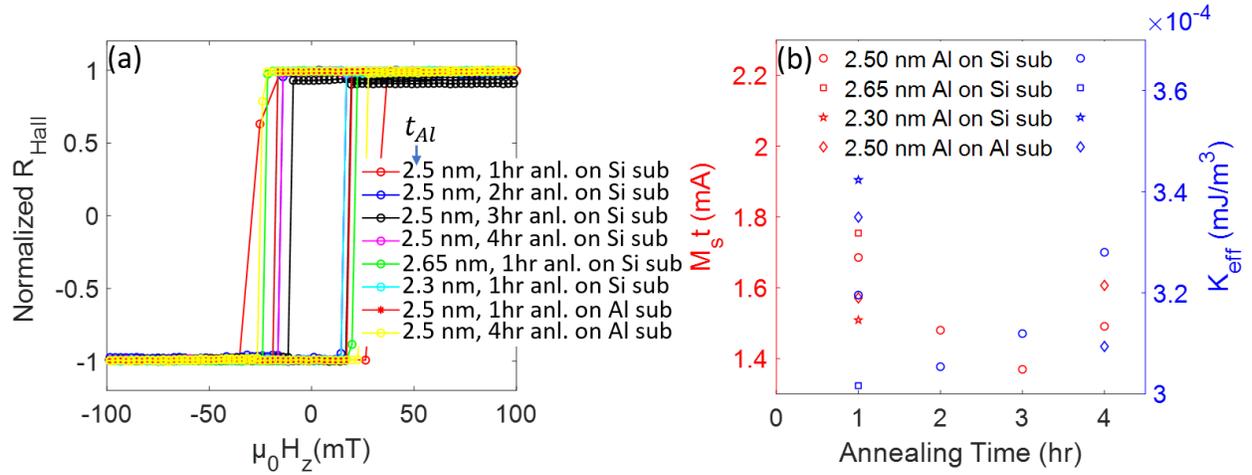

*Fig. 1. (a) Normalized hysteresis loops obtained from anomalous hall effect measurements for the samples varying Al thickness ($t_{Al}$), annealing time, and substrate. (b) Areal magnetization (red) and effective anisotropy energy (blue) as a function of annealing time.*

Next, we utilized ferromagnetic resonance (FMR) spectroscopy to measure the damping parameter and magnetic inhomogeneity in perpendicular geometry, the spectroscopic g-factor, and anisotropy. Our FMR spectrometer consists of a broadband (1-70 GHz) vector network analyzer to apply RF field via a coplanar waveguide (CPW) and measure the S12 parameter under a perpendicular magnetic field up to 2 Tesla. The field homogeneity over the sample volume is better than 0.1% and the field was measured with a Hall probe continuously. The samples are directly placed on the CPW upside-down after coating with a thin polymethyl methacrylate (PMMA) to prevent shorting. The schematic of the sample placement is shown in Fig. 2(c) inset.

Fig. 2 shows examples of measured transmission parameter S12 for sample $Pt(4)/Co(1.2)/AlO_x(2.5)$ as the magnetic field is swept through the resonances. The ferromagnetic resonance can be described by the complex susceptibility $\chi(H_{res})$ obtained from the Landau-Lifshitz equation [35,36]. For the out-of-plane magnetic field geometry, $\chi(H_{res})$ is given by

$$\chi(H_{res}) = \frac{M_{eff}(H_{res}-M_{eff})}{(H_{res}-M_{eff})^2 - H_{eff}^2 - i\Delta H\ (H_{res}-M_{eff})}, \qquad (1)$$

where $H_{res}$ is the resonance field, $\Delta H$ is the linewidth, $H_{eff} = 2\pi f/(\gamma\mu_0)$, $M_{eff} = M_s - \frac{2K_{eff}}{\mu_0 M_s}$ is the effective magnetization, $f$ is the frequency of microwave, $\gamma = g\mu_B/\hbar$ is the gyromagnetic ratio, $\hbar$ is the reduced Planck's constant. Fig. 2(a) and (b) show examples of the spectra of multiple frequencies. We simultaneously fit the real and imaginary parts of Eq. (1) to determine $H_{res}$ and $\Delta H$. The fitted curves are also shown in Fig 2(a) and (b). The resonance field and frequency are related to each other by the Kittle equation

$$f(H_{res}) = \frac{g\mu_0\mu_B}{2\pi\hbar}(H_{res} - M_{eff}). \tag{2}$$

We fit Eq. (2) with the obtained $H_{res}$ fields and show the curves in Fig. 2(c). We find that $M_{eff}$= -0.413 ± 0.01 T and $g$ = 2.144 ± 0.037. The wide error bars associated with the determination of the g-factor may be due to our extremely small magnetic film thickness, high damping, and large linewidth for all frequencies which results in a low signal-to-noise ratio.

In the framework of the Landau-Lifshitz equation, the linewidth linearly depends on the microwave frequency and the phenomenological damping parameter $\alpha$ by

$$\Delta H = \frac{4\pi\alpha}{\gamma\mu_0}f + \Delta H_0, \tag{3}$$

where $\Delta H_0$ is the inhomogeneous broadening. By using Eq. (3), as shown in Fig.S4, we determined $\alpha$=0.169 ± 0.015 and $\mu_0\Delta H_0$= 0.003 ± 0.02 T. Because of the large damping and spin-pumping into the Pt layer, $\Delta H_0$ determination was not precise.

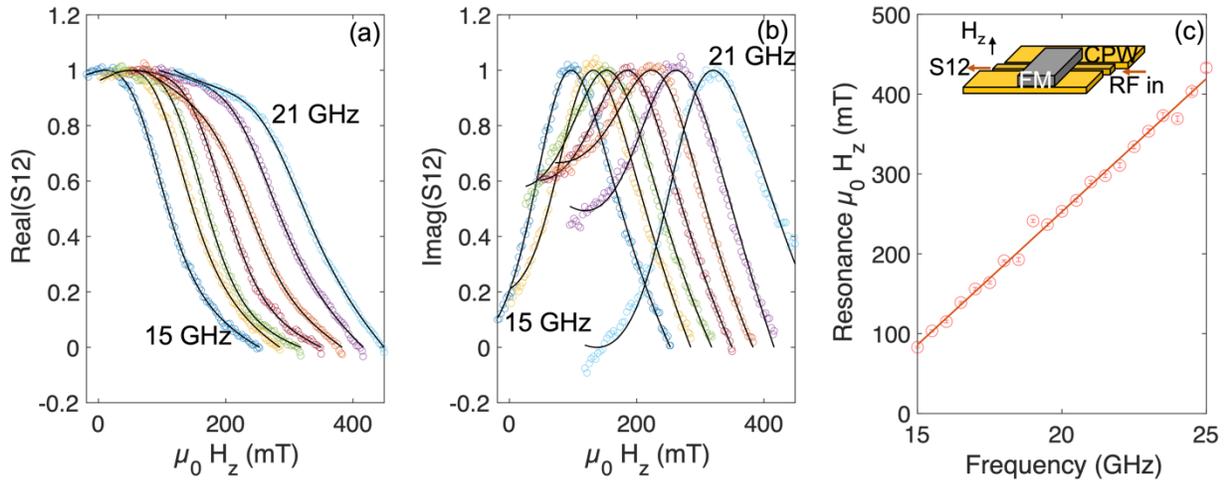

Fig. 2. Ferromagnetic resonance spectrometry results for 1 hour annealed $Pt(4)/Co(1.2)/AlO_x(2.5)$ film. (a) and (b) show the real and imaginary parts of S12 parameter, respectively. (c) shows the resonance field vs microwave frequency and the linear fit.

### b. DMI Characterization

We used polar MOKE microscopy to adopt the method of asymmetric DW expansion in the presence of the in-plane field $(H_x)$ for the quantification of DMI. The DW motion was studied in a creep regime where the velocities of domain walls (DWs) are slow and easy to capture by our MOKE microscopy following millisecond magnetic field pulses. The velocity in this regime depends exponentially on the $H_z$ field [3], i.e., $\ln[v] \propto H_z^{-\frac{1}{4}}$ as shown in Fig.S3(a) of the supplementary material for the sample $Pt(4)/Co(1.2)/Alo_x(2.5)$ annealed for four different time intervals at 350 °C. For the study of DW dynamics in attendance of $H_x$, we saturated our sample by applying a relatively high $\mu_0 H_z = \pm 35\ mT$ fields. Then, the DWs were nucleated by the opposite $H_z$ field pulse which was acting as the reference image for measuring the distance of DWs expansion. During the experiment, the sample was at the center of magnetic pole pieces of the electromagnet to eliminate other stray fields and was aligned horizontally to make sure that there was no $H_z$ component of the in-plane magnetic field as it could influence the DW motion exponentially. Such alignment was tested by applying high $\mu_0 H_x \sim 350\ mT$ in the absence of $H_z$ field and found no expansion of DWs. The reference DWs were expanded in the creep regime by the $0.3\ s\ to\ 2\ s\ H_z$ field pulses under various $H_x$. The DW expansions are symmetric at $\mu_0 H_x = 0\ mT$ as seen in Fig. 3 (a) and (d). However, the right and left side DWs expansion are asymmetric under non-zero $H_x$ field as shown in Fig. 3 (b), (c), (e), and (f). Our explanation for this asymmetry is that the DWs in our magnetic thin films are Neel rather than Bloch type.

In addition, for these thin films which have strong PMA and DMI, the spins at the center of DWs are aligned in a specific direction by the chirality, and an effective magnetic field becomes important referred to as the DMI field ($H_{dmi}$). This field along with $H_x$ has a significant role in the determination of the DW's surface energy and their velocities of expansion [3]. When the DWs expanded due to the $H_z$ pulse, are further subjected to $H_x$ their configuration will be broken, and then down-up (DU) & up-down (UD) DWs acquire different velocities along the direction of $H_x$ as illustrated by the right edge velocity $(v) - H_x$ plots in Fig. 4 (a)-(d). We note that DU (Figure 4 - red data points) and UD (Figure 4 - black data points) DWs have minimum velocities for the specific non-zero value of $H_x$, at which point $H_x$ balances the effective DMI field $H_{dmi}$ [3,29,37]. The extracted $H_{dmi}$ fields are used to calculate DMI constant (D) from the relation; $H_{DMI} = \frac{D}{\mu_0 M_s}\sqrt{K_{eff}/A}$ , where A is the exchange stiffness constant and its value was taken to be $16\ pJ/m$ [3,38]. The $H_{dmi}$ (red data set) and $D$ (blue data set) are plotted as function of annealing time for the different set of samples as given in Fig. 5 (b). Although there are few drawbacks[39,40], we observed a good agreement with the additional DMI measurements on most films.

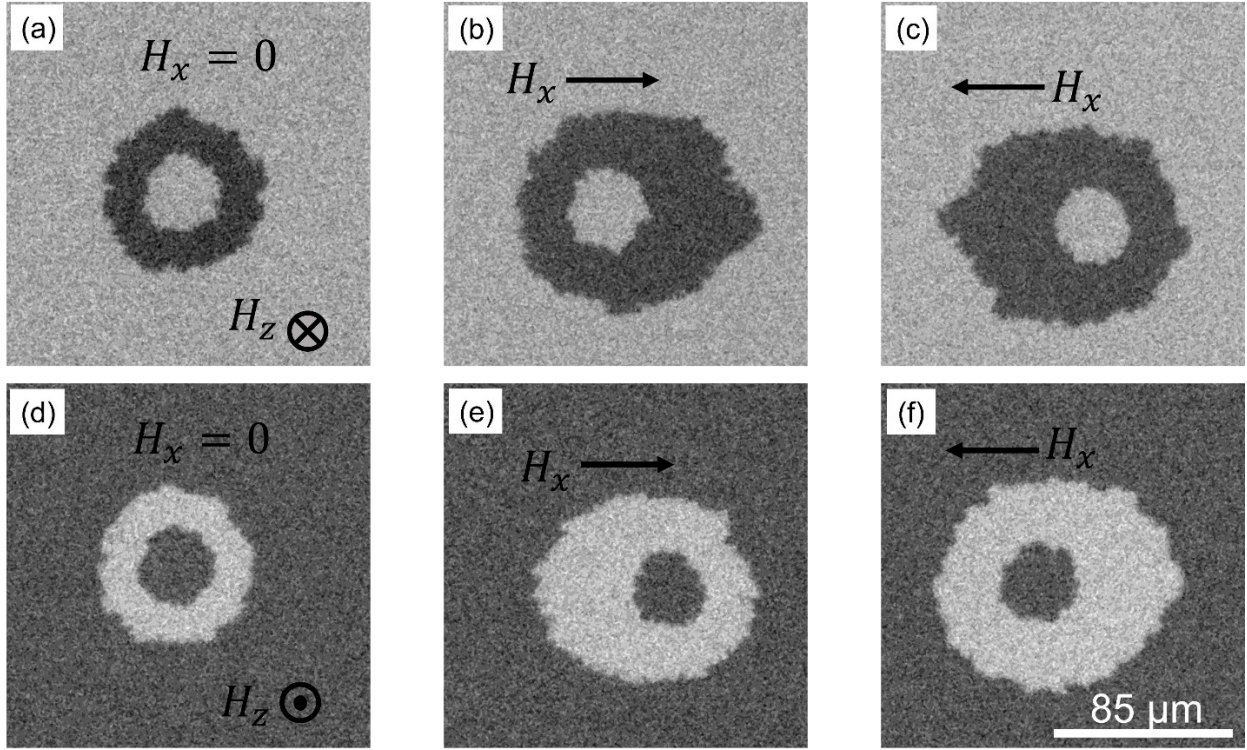

Fig. 3. Expansion of DWs driven by OOP magnetic field $\mu_0 H_z = \pm 9.5\ mT$ observed in Polar MOKE for $Pt(4)/Co(1.2)/AlO_x(2.5)$ grown on $SiO_2/Si$ substrate annealed for 4 hours at various in-plane (IP) magnetic fields. (a),(d) at $\mu_0 H_x = 0\ mT$ with 1 s pulse (b) at $\mu_0 H_x = 85\ mT$ with 2 s pulse (c) at $\mu_0 H_x = -95\ mT$ with 600 ms pulse (e) at $\mu_0 H_x = 75\ mT$ with 2 s pulse (f) at $\mu_0 H_x = -70\ mT$ with 1 s pulse. All images were obtained after subtracting the background images.

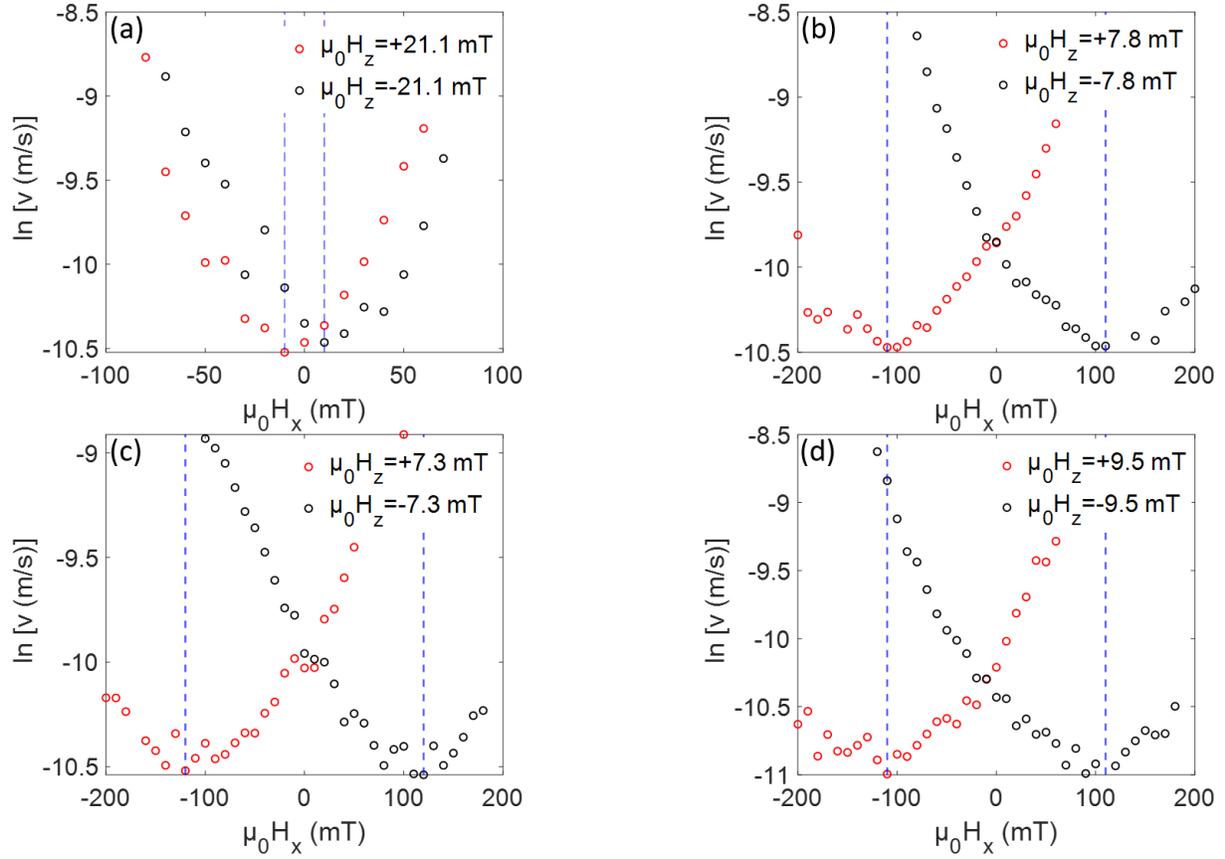

*Fig. 4. Right edge velocity of DWs as a function of in-plane field ($H_x$) for $Pt(4)/Co(1.2)/Alo_x(2.5)$ on a $SiO_2/Si$ substrate with (a) 1-hour (b) 2-hour (c) 3-hour (d) 4-hour annealing.*

Secondly, we adopted a current induced hysteresis loop shift method to quantify the DMI, which was developed by Pai et al. and his co-workers [41]. In this method, when an IP current (I) flows along the x-direction through the heavy metal (i.e. Platinum), the electrons with opposite spins align in the top and bottom interface of thin film, and a spin current is generated along the z-direction. This spin current gives rise to Slonczewski-like SOT, which is responsible for the generation of an effective field ($H_z^{eff}$) along the z-direction as explained by the relation[4][42]: $H_z^{eff} \propto \boldsymbol{m} \times (\boldsymbol{z} \times \boldsymbol{J_x})$ where $J_x$ is the magnitude of charge current density along x-direction. This effective field is equal and opposite to the DMI caused by DU and UD Neel DWs because of their antiparallel orientation of magnetic moments. However, in presence of $H_x$, the moments of chiral DWs tend to align along the direction $H_x$ that produces an opposite $H_z^{eff}$ with an unequal magnitude. So there exist net positive (or negative) $H_z^{eff}$ field depending on the direction of $H_x$ and in-plane DC current (I). This net effective field is responsible for shifting of the hysteresis loop as seen in the Fig. 5 (b) and the measured shift can be taken as $H_z^{eff}$. In this plot, black data points and red data points represent the anomalous Hall resistance measurement as a function of $H_z$ field for positive (+15 mA) and negative (-15 mA) IP currents respectively with the bias field of 146 mT for $Pt(4)/Co(1.2)/AlO_x(2.5)$ sample annealed for three hours. Moreover, $H_z^{eff}$ depends linearly with the current density as verified by the plot of $H_z^{eff}$ as a function of I in presence of two equal and opposite IP biased fields given in Fig. 5 (c). While calculating loop shift, $H_z^{eff}$, we were aware that DWs switching field could change due to IP current as well as Joule's heating effect. The latter effect was eliminated by measuring shift of center of hysteresis which was achieved by taking the average of the absolute values of switching fields for DU ($H_{D-U}^{sw}$) and UD ($H_{U-D}^{sw}$) DWs[41,43]. Next, we measured the shift for several $H_x$ and found it saturates as $H_x$ increases, as the magnetizations of the DMI-induced Neel DWs aligned in the direction $H_x$. This specific value of $H_x$ was taken as $H_{dmi}$ field as shown in the plot in Fig. 5 (d) and the hall bar devices for the measurement of DMI is given in the Fig. 5 (a). We found $\mu_0 H_{dmi}$ = 131 and 123 ($\pm$ 5) $mT$ for the $Pt(4)/Co(1.2)/Alo_x(2.5)$ samples annealed for 3 and 4 hours, respectively. By using the same equation as used by the asymmetric domain wall expansion section, the D values were calculated to be $1.13 \pm 0.04\ mJ/m^2$ and $1.06 \pm 0.03\ mJ/m^2$ for the respective samples.

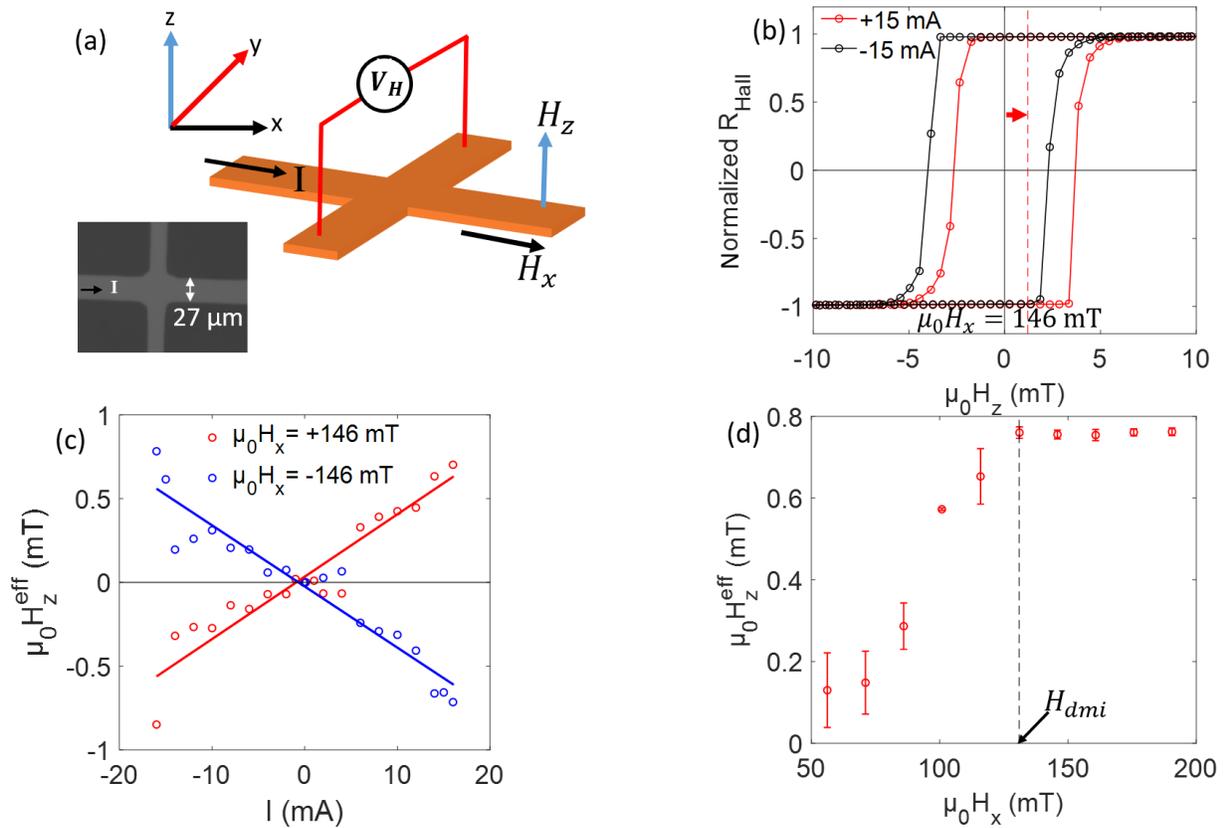

*Fig. 5. (a) Schematic for the hysteresis-shift DMI measurement for the under the 2-dimensional magnetic fields and picture of Hall bar device with 27 μm channel width. (b) Normalized Hall resistance measurement in presence of bias field 146 mT with DC of +15 mA (red) and -15 mA (black) along the x-direction. (c) Effective $H_z$ field as a function of in-plane DC for the sample $Pt(4)/Co(1.2)/AlO_x(2.5)$ on SiO₂/Si substrate annealed for 3 hours; blue and red symbols represent $H_z^{eff}$ in the presence of positive and negative fields respectively. (d) Effective Hall field along z-direction due to spin hall effect as a function of IP field.*

Finally, DMI of $Pt(4)/Co(1.2)/AlO_x(2.5)$ stacks were studied using Brillouin Light Spectroscopy (BLS), where spin-wave (thermal magnons) frequencies are measured in ferromagnetic materials. In MLs with broken inversion symmetry at the interfaces, the DMI modifies the spin-wave dispersion relation and dispersion curves become asymmetric, known as non-reciprocal spin-wave dispersion, as illustrated by Fig. 6 (a) and (b). The frequency shift ($\Delta f$) due to the DMI is given by [44]:

$\Delta f = \frac{2g\,\mu\,D\,k_x}{h\,M_s}$ where g, $\mu, h, M_s, k_x$ are an in-plane spectroscopic splitting factor, Bohr magneton, Planck's constant, saturation magnetization, and the magnitude of spin waves vector along x-direction respectively. The detailed experimental techniques can be found in the reference [44]. The $\Delta f$ of $Pt/Co/AlO_x(2.5\,nm)/Pt$ stack for 1 hour and 4 hour annealed samples were measured to be $5.7 \times 10^{-1} \pm 2 \times 10^{-2}$ GHz and $4.9 \times 10^{-1} \pm 2 \times 10^{-2}$ GHz respectively. The corresponding DMI values calculated by using the above equation are $0.82 \pm 0.03\,mJ/m^2$ and $0.62 \pm 0.03\,mJ/m^2$ respectively. We note that the sample annealed for four hours shows larger full width at half maximum (FWHM) compared to the one annealed for one hour. The discrepancy between the FWHMs of Fig. 6 (a) and (b) curves can be attributed to the differences in surface roughness, damping, or spin pumping which can alter during a longer annealing process.

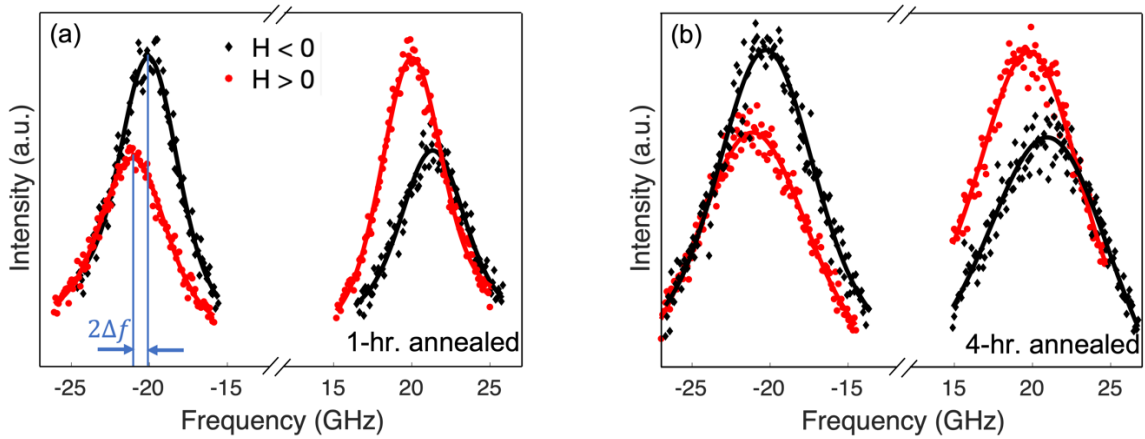

*Fig. 6. Spin wave dispersion curves obtained from BLS measurement technique for (a) 1-hour (b) 4-hour annealed $Pt(4)/Co(1.2)/Alo_x(2.5)$ sample. The black and red data points spectra were measured under the presence of a magnetic field with two opposite polarities.*

To summarize this section, the DMI constants (D) were accentuated up to around $1.13\,mJ/m^2$ for $Pt(4)/Co(1.2)/AlO_x(t)$ as agreed by all three measurement techniques as displayed in Fig. 7 though there are tiny deviations in their quantitatively measured values. The minimum value of D was recorded as low as $0.1\,mJ/m^2$ from the study of field-driven DW dynamics. However, this D value is weaker than that of the BLS measurement result for the same sample. It might be due to the different approaches of probing the film in BLS and p-MOKE microscopy measurement techniques [30,45,46]. Initially, that non-zero D is attributed to the symmetry breaking when the $AlO_x$, which has very weak SOC and no contribution on the DMI on its own, is introduced on the symmetric $Pt/Co/Pt$. In addition to that, several other interesting mechanisms are responsible for the variation of D due to the insertion of $AlO_x$. First, the Co layer might be partially oxidized as it is in direct contact with $AlO_x$. The effective Co thickness is then reduced, which changes the DMI with the $Pt$ layer[31,47]. Second, the annealing time can alter the epitaxial relationship between the layers and the bottom $Pt$ layer. This might induce lattice strain at the interface, where the SOC is affected due to the lattice mismatch between top $Co$ and $AlO_x$ layers [48]. Moreover, a longer thermal annealing duration might improve the ordering of atoms at the Pt/Co interface giving rise to DMI, since the DMI is susceptible to the atomic arrangement at the interfaces [49,50]. Therefore, the bottom $Pt/Co$ interface may also be partially responsible for the alteration of interfacial DMI. Finally, a large interfacial electric field induced by the interfacial oxidation, originating from the transfer of charge, surpasses the very small SOC [51] of the atoms at the $Co/AlO_x$ interface contribute for the large DMI [52]. In our case, the degree of oxidation of cobalt is changed as it was annealed for distinct time intervals leaving less cobalt in between $Pt$ and $AlO_x$ layers as confirmed by the XPS results that would be discussed in the material characterization section in detail. Thus, the enhancement in the DMI for the tri-layer system grown on the same substrate is due to the interface modification of the top $Co/AlO_x$ interface. On the other hand, for our same stacks grown on two different substrates (i.e. SiO$_2$/Si and Al$_2$O$_3$), the DMI values differ by some magnitude that might be due to the different lattice mismatch between magnetic films and the substrates. Those lattice mismatches induce the discrete lattice strains that alters the anisotropy and DMI as well [53–55].

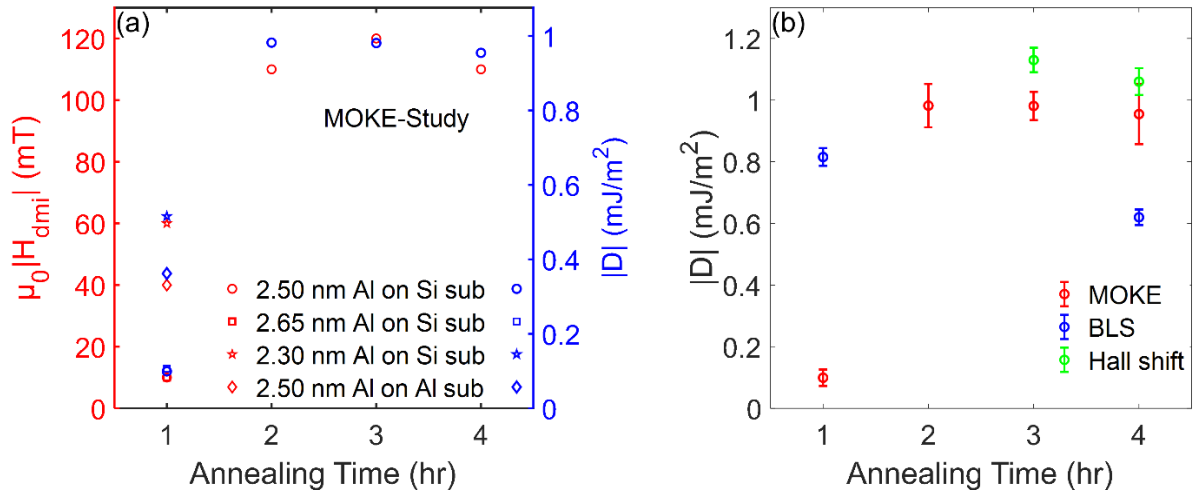

*Fig. 7 (a) DMI field (red) and DMI constant (blue) as a function of annealing time; circles, squares, stars and diamonds symbols represent $Pt(4)/Co(1.2)/AlO_x(2.5)$, $Pt(4)/Co(1.2)/AlO_x(2.65)$, $Pt(4)/Co(1.2)/AlO_x(2.3)$ grown on $SiO_2/Si$ substrate and $Pt(4)/Co(1.2)/AlO_x(2.5)$ grown on Al$_2$O$_3$ substrate, respectively (b) D constants as a function of annealing time quantified by using MOKE microscopy (red data points), BLS (blue data points) and Hall shift measurement (green data points) techniques for the sample $Pt(4)/Co(1.2)/AlO_x(2.5)$.*

### c. Material characterization

Following our magnetic characterization, we employed X-ray photoelectron spectroscopy (XPS) with $Ar^+$ etching technique for the investigation of the chemical composition of our multilayer stack. The high-resolution spectra of Co 2p core-level state, from as-grown to 90 minutes etching time in every 15 minutes time interval, for 1 hour and 3 hours annealed $Pt(4)/Co(1.2)/AlO_x(2.5)$ heterostructures are included in the supplementary materials Fig. S2(a) and (b). All the as-grown samples do not show the signal of Co, this is because all samples have a protective thick Pt layer on top of it, but once this layer has been removed by $Ar^+$ sputtering, the signal from cobalt starts to appear. Particularly, a peak centered around binding energy of 778.1 eV is clearly seen. This peak indicates the presence of metallic cobalt underneath the Pt capping layer.

To study the oxidation state of the Co at the $Co/AlO_x$ interface, we have analyzed the spectra (Fig 8) from the 75 minutes sputtered samples for the same $Pt(4)/Co(1.2)/AlO_x(2.5)$ stack. In the plots, the high-resolution XPS spectra of the Co 2p region, for samples annealed for 1 hour (Fig. 8a) and 3 hours (Fig. 8b), have been deconvoluted using Gaussian fitting. Both spectra can be deconvoluted into four different sub-peaks: Metallic cobalt, $Co^{2+}$, $Co^{3+}$, and $Co$ satellite. These peaks in both samples are centered at binding energies of 778.3 eV, 780.3 eV, 782.4 eV, and 785.0 eV, respectively. These results confirm the co-existence of $Co$ and $CoO_x$ at the interface [43,56]. However, the area under these peaks varies for each sample, which is an indication of different concentrations of Co states depending on the annealing time. The estimated ratios of metallic $Co$ and the oxide of $Co$ were found to be 1:0.89 and 1:1.35 with the annealing times 1 hour and 3 hours, respectively, which is evidence for the significant difference in oxygen content at the $Co - Al$ interface of the discrete time-annealed sample. In addition, we performed the low temperature anomalous Hall effect (LT-AHE) measurement technique as presented in Fig 8(c) and more detailed plots in Fig.S3(b) of the supplementary material. We extracted the coercivities at the measured temperatures ranging from 15 K to 300 K. All of the four samples have relatively larger coercivities at low temperatures and $H_c$ is highest for the 3 hours annealed sample confirming the strong oxidation states due to the oxygen migration at the interface [57]. However, on increasing the annealing time further (i.e. for the four-hour annealed sample) the coercivity value began to reduce which might be attributed to the diffusion of oxygen away from the cobalt layer [25,26]. In addition, a cross-sectional view of our multilayer magnetic sample using transmission electron microscopy is displayed in Fig. 8 (d). The individual layers in our samples are easily distinguished and show clear modification at the Co and Al interface due to the annealing-induced diffusion of oxygen.

Both XPS and LT-AHE results showed a very good agreement about the oxide concentration on the Co and Al interface and found that it was highest for the three-hour annealed sample. This physical characterization supports our claim about the impact of oxygen content on the interface for tuning the DMI as calculated and stated in the previous section for the $Pt/Co/AlO_x$ tri-layers for different annealing time intervals and Al thicknesses were grown on the silicon wafer.

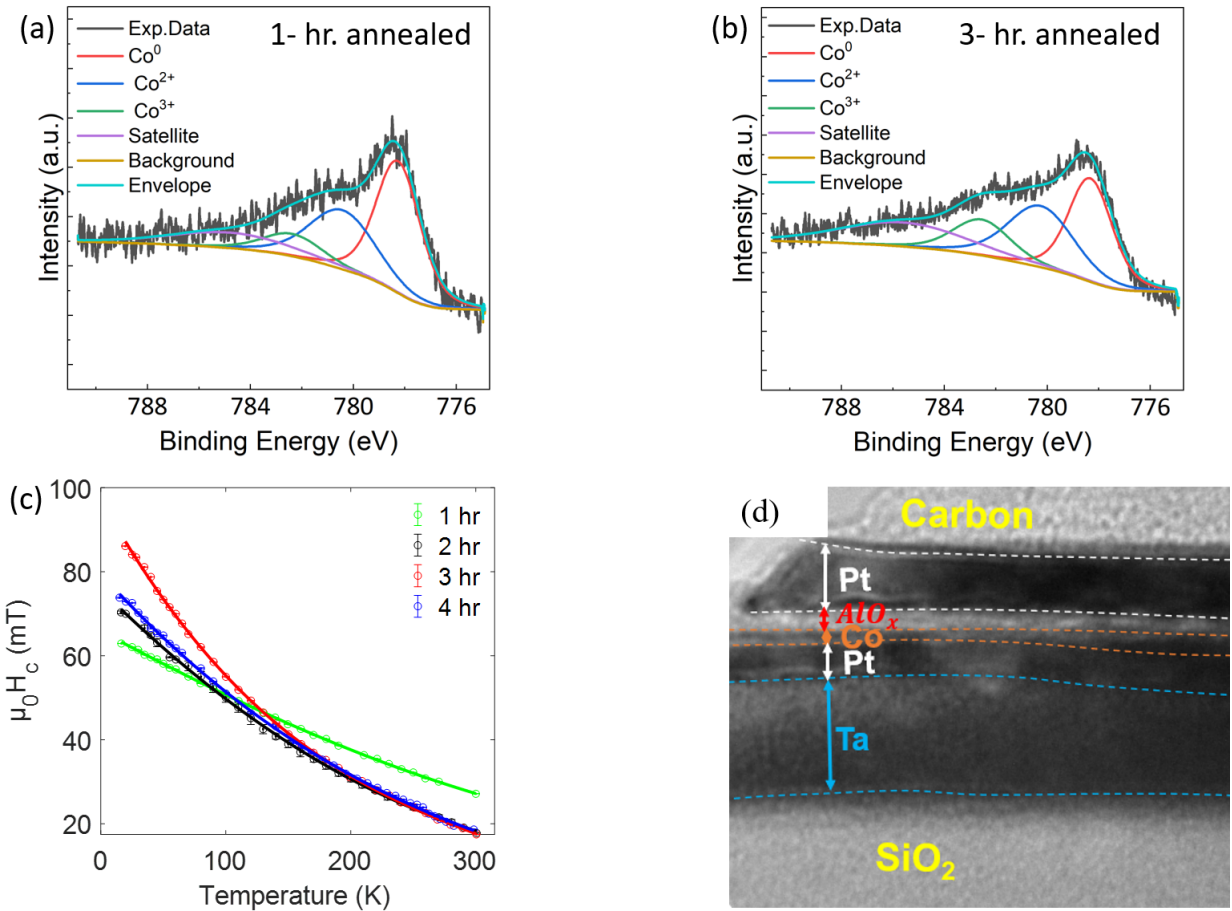

*Fig. 8. (a) and (b)The experimental and fitted XPS spectra of Co 2p state for $Pt(4)/Co(1.2)/Alo_x(2.5)$ annealed for 1 hour and 3 hours respectively; (c) Coercive fields extracted from the hysteresis curves of low-temperature Hall measurement techniques Vs temperature. (d) TEM images for the same sample annealed at 1 hour.*

## 4. Conclusion

In summary, we have shown that variation of the oxygen level at the cobalt-aluminum interface in a perpendicularly magnetized $Pt(2)/Co(1.2)/AlO_x(t)$ trilayer structure can significantly affect the interfacial Dzyaloshinskii-Moria interaction. We measured the enhanced DMI constant from $0.10 \pm 0.02\ mJ/m^2$ to $1.13 \pm 0.04\ mJ/m^2$ using comprehensive characterization methods including MOKE microscopy, Brillouin light spectroscopy, and hysteresis loop shift measurement technique for samples with different annealing times, Aluminum thickness, and the base substrate. Interestingly, we found little deviation in the values obtained from each approach for the specific samples, with the trend of incremental changes in the DMI remaining the same. This is due to the variation of oxygen content at the interface as illustrated by the results obtained from X-ray photoelectron microscopy, and low-temperature hall effect measurement techniques. Our study is helpful for the optimization of material combination for designing ultrathin film for the stabilization of skyrmions by enhancing the DMI desired for use in Domain wall and racetrack memories.

ACKNOWLEDGEMENT

This work is supported by NSF OIA-1929086.

References:

[1] I. DZYALOSHINSKY, A THERMODYNAMIC THEORY OF "WEAK" FERROMAGNETISM OF ANTIFERROMAGNETICS, J. Phys. Chem. Solids Pergamon Press. 4 (1958) 241–255. doi:10.1002/pssb.2220460236.

[2] T. Moriya, New mechanism of anisotropic superexchange interaction, Phys. Rev. Lett. 4 (1960) 228–230. doi:10.1103/PhysRevLett.4.228.

[3] D.K. Lau, B.S. Nanoengineering, Experimental Evaluation of the Interfacial Dzyaloshinskii-Moriya Interaction in Co/Ni Magnetic Multilayers, (2018).

[4] S. Emori, U. Bauer, S.M. Ahn, E. Martinez, G.S.D. Beach, Current-driven dynamics of chiral ferromagnetic domain walls, Nat. Mater. 12 (2013) 611–616. doi:10.1038/nmat3675.

[5] J. Sampaio, V. Cros, S. Rohart, A. Thiaville, A. Fert, Nucleation, stability and current-induced motion of isolated magnetic skyrmions in nanostructures, Nat. Nanotechnol. 8 (2013) 839–844. doi:10.1038/nnano.2013.210.

[6] G. Siracusano, R. Tomasello, A. Giordano, V. Puliafito, B. Azzerboni, O. Ozatay, M. Carpentieri, G. Finocchio, Magnetic Radial Vortex Stabilization and Efficient Manipulation Driven by the Dzyaloshinskii-Moriya Interaction and Spin-Transfer Torque, Phys. Rev. Lett. (2016). doi:10.1103/PhysRevLett.117.087204.

[7] A.N. Bogdanov, U.B. Rößler, Chiral symmetry breaking in magnetic thin films and multilayers, Phys. Rev. Lett. 87 (2001) 37203-1-37203–4. doi:10.1103/PhysRevLett.87.037203.

[8] J. Iwasaki, M. Mochizuki, N. Nagaosa, Current-induced skyrmion dynamics in constricted geometries, Nat. Nanotechnol. 8 (2013) 742–747. doi:10.1038/nnano.2013.176.

[9] R. Juge, S.G. Je, D.D.S. Chaves, L.D. Buda-Prejbeanu, J. Peña-Garcia, J. Nath, I.M. Miron, K.G. Rana, L. Aballe, M. Foerster, F. Genuzio, T.O. Menteş, A. Locatelli, F. Maccherozzi, S.S. Dhesi, M. Belmeguenai, Y. Roussigné, S. Auffret, S. Pizzini, G. Gaudin, J. Vogel, O. Boulle, Current-Driven Skyrmion Dynamics and Drive-Dependent Skyrmion Hall Effect in an Ultrathin Film, Phys. Rev. Appl. 12 (2019) 1–9. doi:10.1103/PhysRevApplied.12.044007.

[10] A. Fert, V. Cros, J. Sampaio, Skyrmions on the track, Nat. Nanotechnol. 8 (2013) 152–156. doi:10.1038/nnano.2013.29.

[11] L.T. S.S.P.Parkin,M.Hayashi, Magnetic Domain-Wall Racetrack Memories, J. Chem. Inf. Model. 320 (2008).

[12] Y. Wu, K. Meng, J. Miao, X. Xu, Y. Jiang, Enhanced spin-orbit torque in Pt/Co/Pt multilayers with inserting Ru layers, J. Magn. Magn. Mater. 472 (2019) 14–19. doi:10.1016/j.jmmm.2018.10.012.

[13] D. Claudio-Gonzalez, A. Thiaville, J. Miltat, Domain wall dynamics under nonlocal spin-transfer

torque, Phys. Rev. Lett. (2012). doi:10.1103/PhysRevLett.108.227208.

[14] S. Zhang, Z. Li, Roles of nonequilibrium conduction electrons on the magnetization dynamics of ferromagnets, Phys. Rev. Lett. (2004). doi:10.1103/PhysRevLett.93.127204.

[15] D. Li, B. Cui, J. Yun, M. Chen, X. Guo, K. Wu, X. Zhang, Y. Wang, J. Mao, Y. Zuo, J. Wang, L. Xi, Current-Induced Domain Wall Motion and Tilting in Perpendicularly Magnetized Racetracks, Nanoscale Res. Lett. 13 (2018). doi:10.1186/s11671-018-2655-6.

[16] L. Liu, C.F. Pai, Y. Li, H.W. Tseng, D.C. Ralph, R.A. Buhrman, Spin-torque switching with the giant spin hall effect of tantalum, Science (80-. ). 336 (2012) 555–558. doi:10.1126/science.1218197.

[17] J.C. Martinez, W.S. Lew, W.L. Gan, M.B.A. Jalil, Theory of current-induced skyrmion dynamics close to a boundary, J. Magn. Magn. Mater. 465 (2018) 685–691. doi:10.1016/j.jmmm.2018.06.031.

[18] W. Jiang, X. Zhang, G. Yu, W. Zhang, X. Wang, M. Benjamin Jungfleisch, J.E. Pearson, X. Cheng, O. Heinonen, K.L. Wang, Y. Zhou, A. Hoffmann, S.G.E. Te Velthuis, Direct observation of the skyrmion Hall effect, Nat. Phys. 13 (2017) 162–169. doi:10.1038/nphys3883.

[19] N. Romming, A. Kubetzka, C. Hanneken, K. Von Bergmann, R. Wiesendanger, Field-dependent size and shape of single magnetic Skyrmions, Phys. Rev. Lett. (2015). doi:10.1103/PhysRevLett.114.177203.

[20] Y. Chen, Q. Zhang, J. Jia, Y. Zheng, Y. Wang, X. Fan, J. Cao, Tuning Slonczewski-like torque and Dzyaloshinskii-Moriya interaction by inserting a Pt spacer layer in Ta/CoFeB/MgO structures, Appl. Phys. Lett. (2018). doi:10.1063/1.5026423.

[21] J. Torrejon, J. Kim, J. Sinha, S. Mitani, M. Hayashi, M. Yamanouchi, H. Ohno, Interface control of the magnetic chirality in CoFeB/MgO heterostructures with heavy-metal underlayers, Nat. Commun. 5 (2014) 4–11. doi:10.1038/ncomms5655.

[22] M. Kuepferling, A. Casiraghi, G. Soares, G. Durin, F. Garcia-Sanchez, L. Chen, C.H. Back, C.H. Marrows, S. Tacchi, G. Carlotti, Measuring interfacial dzyaloshinskii-moriya interaction in ultra thin films, ArXiv. (2020).

[23] H.T. Nembach, E. Jué, E.R. Evarts, J.M. Shaw, Correlation between Dzyaloshinskii-Moriya interaction and orbital angular momentum at an oxide-ferromagnet interface, Phys. Rev. B. (2020). doi:10.1103/PhysRevB.101.020409.

[24] M. Arora, J.M. Shaw, H.T. Nembach, Variation of sign and magnitude of the Dzyaloshinskii-Moriya interaction of a ferromagnet with an oxide interface, Phys. Rev. B. (2020). doi:10.1103/PhysRevB.101.054421.

[25] B. Rodmacq, A. Manchon, C. Ducruet, S. Auffret, B. Dieny, Influence of thermal annealing on the perpendicular magnetic anisotropy of Pt/Co/AlOx trilayers, Phys. Rev. B - Condens. Matter Mater. Phys. 79 (2009) 1–8. doi:10.1103/PhysRevB.79.024423.

[26] A. Manchon, S. Pizzini, J. Vogel, V. Uhlíř, L. Lombard, C. Ducruet, S. Auffret, B. Rodmacq, B. Dieny, M. Hochstrasser, G. Panaccione, X-ray analysis of oxygen-induced perpendicular magnetic anisotropy in Pt / Co / AlOx trilayers, J. Magn. Magn. Mater. (2008). doi:10.1016/j.jmmm.2008.02.131.

[27] H.K. Gweon, S.H. Lim, Relative strength of perpendicular magnetic anisotropy at bottom and top interfaces in Pt/Co/MgO trilayers, Jpn. J. Appl. Phys. 57 (2018). doi:10.7567/JJAP.57.030301.

[28] A. Hrabec, N.A. Porter, A. Wells, M.J. Benitez, G. Burnell, S. McVitie, D. McGrouther, T.A. Moore, C.H. Marrows, Measuring and tailoring the Dzyaloshinskii-Moriya interaction in perpendicularly magnetized thin films, Phys. Rev. B - Condens. Matter Mater. Phys. 90 (2014) 1–5. doi:10.1103/PhysRevB.90.020402.

[29] A. Cao, R. Chen, X. Wang, X. Zhang, S. Lu, S. Yan, B. Koopmans, W. Zhao, Enhanced interfacial Dzyaloshinskii - Moriya interactions in annealed Pt/Co/MgO structures, Nanotechnology. 31 (2020). doi:10.1088/1361-6528/ab62cd.

[30] R. Soucaille, M. Belmeguenai, J. Torrejon, J. V. Kim, T. Devolder, Y. Roussigné, S.M. Chérif, A.A. Stashkevich, M. Hayashi, J.P. Adam, Probing the Dzyaloshinskii-Moriya interaction in CoFeB ultrathin films using domain wall creep and Brillouin light spectroscopy, Phys. Rev. B. 94 (2016) 1–8. doi:10.1103/PhysRevB.94.104431.

[31] A. Cao, X. Zhang, B. Koopmans, S. Peng, Y. Zhang, Z. Wang, S. Yan, H. Yang, W. Zhao, Tuning the Dzyaloshinskii-Moriya interaction in Pt/Co/MgO heterostructures through the MgO thickness, Nanoscale. 10 (2018) 12062–12067. doi:10.1039/c7nr08085a.

[32] B.R. Sankhi, E. Turgut, A low-cost vibrating sample magnetometry based on audio components, J. Magn. Magn. Mater. 502 (2020) 166560. doi:10.1016/j.jmmm.2020.166560.

[33] Johnson M. T., Bloemen P. J. H., den Broeder F. J. A., J.J. de Vries, Magnetic anisotropy in metallic multilayers, Reports Prog. Phys. (1996).

[34] K. Yakushiji, H. Kubota, A. Fukushima, S. Yuasa, Perpendicular magnetic tunnel junction with enhanced anisotropy obtained by utilizing an Ir/Co interface, Appl. Phys. Express. (2016). doi:10.7567/APEX.9.013003.

[35] J.M. Shaw, H.T. Nembach, T.J. Silva, C.T. Boone, Precise determination of the spectroscopic g-factor by use of broadband ferromagnetic resonance spectroscopy, J. Appl. Phys. (2013). doi:10.1063/1.4852415.

[36] J.M. Shaw; H.T. Nembach; T.J. Silva, Measurement of orbital asymmetry and strain in Co90Fe10/Ni multilayers and alloys: Origins of perpendicular anisotropy.pdf, (2013). doi:10.1103/PhysRevB.87.054416.

[37] A. Cao, R. Chen, X. Wang, X. Zhang, S. Lu, S. Yan, B. Koopmans, W. Zhao, Enhanced interfacial Dzyaloshinskii - Moriya interactions in annealed Pt/Co/MgO structures, Nanotechnology. 31 (2020). doi:10.1088/1361-6528/ab62cd.

[38] S.G. Je, D.H. Kim, S.C. Yoo, B.C. Min, K.J. Lee, S.B. Choe, Asymmetric magnetic domain-wall motion by the Dzyaloshinskii-Moriya interaction, Phys. Rev. B - Condens. Matter Mater. Phys. 88 (2013) 1–5. doi:10.1103/PhysRevB.88.214401.

[39] M. Vaňatka, J.C. Rojas-Sánchez, J. Vogel, M. Bonfim, M. Belmeguenai, Y. Roussigné, A. Stashkevich, A. Thiaville, S. Pizzini, Velocity asymmetry of Dzyaloshinskii domain walls in the creep and flow regimes, J. Phys. Condens. Matter. (2015). doi:10.1088/0953-8984/27/32/326002.

[40] A. Thiaville, S. Rohart, É. Jué, V. Cros, A. Fert, Dynamics of Dzyaloshinskii domain walls in ultrathin

magnetic films, Epl. 100 (2012). doi:10.1209/0295-5075/100/57002.

[41] C.F. Pai, M. Mann, A.J. Tan, G.S.D. Beach, Determination of spin torque efficiencies in heterostructures with perpendicular magnetic anisotropy, Phys. Rev. B. 93 (2016) 1–7. doi:10.1103/PhysRevB.93.144409.

[42] D. Khadka, S. Karayev, S.X. Huang, Dzyaloshinskii-Moriya interaction in Pt/Co/Ir and Pt/Co/Ru multilayer films, J. Appl. Phys. 123 (2018). doi:10.1063/1.5021090.

[43] D. Li, R. Ma, B. Cui, J. Yun, Z. Quan, Y. Zuo, L. Xi, X. Xu, Effect of the oxide layer on the interfacial Dyzaloshinskii-Moriya interaction in perpendicularly magnetized Pt/Co/SmOx and Pt/Co/AlOx heterostructures, Appl. Surf. Sci. 513 (2020). doi:10.1016/j.apsusc.2020.145768.

[44] H.T. Nembach, J.M. Shaw, M. Weiler, E. Jué, T.J. Silva, Linear relation between Heisenberg exchange and interfacial Dzyaloshinskii-Moriya interaction in metal films, Nat. Phys. 11 (2015) 825–829. doi:10.1038/nphys3418.

[45] K. Shahbazi, J. Von Kim, H.T. Nembach, J.M. Shaw, A. Bischof, M.D. Rossell, V. Jeudy, T.A. Moore, C.H. Marrows, Domain-wall motion and interfacial Dzyaloshinskii-Moriya interactions in Pt/Co/Ir(tIr)/Ta multilayers, Phys. Rev. B. (2019). doi:10.1103/PhysRevB.99.094409.

[46] K. Zeissler, M. Mruczkiewicz, S. Finizio, J. Raabe, P.M. Shepley, A. V. Sadovnikov, S.A. Nikitov, K. Fallon, S. McFadzean, S. McVitie, T.A. Moore, G. Burnell, C.H. Marrows, Pinning and hysteresis in the field dependent diameter evolution of skyrmions in Pt/Co/Ir superlattice stacks, Sci. Rep. (2017). doi:10.1038/s41598-017-15262-3.

[47] R. Lo Conte, G. V. Karnad, E. Martinez, K. Lee, N.H. Kim, D.S. Han, J.S. Kim, S. Prenzel, T. Schulz, C.Y. You, H.J.M. Swagten, M. Kläui, Ferromagnetic layer thickness dependence of the Dzyaloshinskii-Moriya interaction and spin-orbit torques in Pt\Co\AlOx, AIP Adv. (2017). doi:10.1063/1.4990694.

[48] P. V. Ong, N. Kioussis, P.K. Amiri, K.L. Wang, G.P. Carman, Strain control magnetocrystalline anisotropy of Ta/FeCo/MgO heterostructures, J. Appl. Phys. (2015). doi:10.1063/1.4916115.

[49] R. Lavrijsen, D.M.F. Hartmann, A. Van Den Brink, Y. Yin, B. Barcones, R.A. Duine, M.A. Verheijen, H.J.M. Swagten, B. Koopmans, Asymmetric magnetic bubble expansion under in-plane field in Pt/Co/Pt: Effect of interface engineering, Phys. Rev. B - Condens. Matter Mater. Phys. (2015). doi:10.1103/PhysRevB.91.104414.

[50] A. Cao, R. Chen, X. Wang, X. Zhang, S. Lu, S. Yan, B. Koopmans, W. Zhao, Enhanced interfacial Dzyaloshinskii - Moriya interactions in annealed Pt/Co/MgO structures, Nanotechnology. (2020). doi:10.1088/1361-6528/ab62cd.

[51] N.H. Kim, J. Cho, J. Jung, D.S. Han, Y. Yin, J.S. Kim, H.J.M. Swagten, K. Lee, M.H. Jung, C.Y. You, Role of top and bottom interfaces of a Pt/Co/AlOx system in Dzyaloshinskii-Moriya interaction, interface perpendicular magnetic anisotropy, and magneto-optical Kerr effect, AIP Adv. 7 (2017). doi:10.1063/1.4978867.

[52] A. Belabbes, G. Bihlmayer, S. Blügel, A. Manchon, Oxygen-enabled control of Dzyaloshinskii-Moriya Interaction in ultra-thin magnetic films, Sci. Rep. (2016). doi:10.1038/srep24634.

[53] N.S. Gusev, A. V. Sadovnikov, S.A. Nikitov, M. V. Sapozhnikov, O.G. Udalov, Manipulation of the Dzyaloshinskii-Moriya Interaction in Co/Pt Multilayers with Strain, Phys. Rev. Lett. 124 (2020)

157202. doi:10.1103/PhysRevLett.124.157202.

[54] C. Deger, Strain-enhanced Dzyaloshinskii–Moriya interaction at Co/Pt interfaces, Sci. Rep. (2020). doi:10.1038/s41598-020-69360-w.

[55] C.J. Pan, T.H. Gao, N. Itogawa, T. Harumoto, Z.J. Zhang, Y. Nakamura, J. Shi, Large lattice mismatch induced perpendicular magnetic anisotropy and perpendicular exchange bias in CoPt/FeMn bilayer films, Sci. China Technol. Sci. (2019). doi:10.1007/s11431-019-1433-0.

[56] M.C. Biesinger, B.P. Payne, A.P. Grosvenor, L.W.M. Lau, A.R. Gerson, R.S.C. Smart, Resolving surface chemical states in XPS analysis of first row transition metals, oxides and hydroxides: Cr, Mn, Fe, Co and Ni, Appl. Surf. Sci. (2011). doi:10.1016/j.apsusc.2010.10.051.

[57] H. Garad, F. Fettar, F. Gay, Y. Joly, S. Auffret, B. Rodmacq, B. Dieny, L. Ortega, Temperature Variation of Magnetic Anisotropy in Pt /Co /AlOx Trilayers, Phys. Rev. Appl. (2017). doi:10.1103/PhysRevApplied.7.034023.